\documentclass[11pt]{article}
\usepackage[T1]{fontenc}
\usepackage{lmodern}
\usepackage[margin=1in]{geometry}
\usepackage{amsmath,amssymb,amsthm,booktabs,microtype}
\usepackage[hidelinks]{hyperref}
\usepackage{bbold}
\usepackage{authblk}

\hypersetup{pdftitle={The covering radius of RM(1,9) is precisely 242}}
\newtheorem{theorem}{Theorem}
\newtheorem{lemma}[theorem]{Lemma}
\newtheorem{proposition}[theorem]{Proposition}
\newcommand{\F}{\mathbb F_2}
\newcommand{\V}{\F^9}
\newcommand{\RM}{\operatorname{RM}}
\newcommand{\NL}{\operatorname{NL}}
\newcommand{\wt}{\operatorname{wt}}
\newcommand{\rank}{\operatorname{rank}}
\newcommand{\sgn}{\operatorname{sgn}}
\newcommand{\GL}{\operatorname{GL}}
\newcommand{\supp}{\operatorname{supp}}
\newcommand{\M}[1]{\left\lVert #1\right\rVert_\infty}
\newcommand{\one}{\mathbb{1}}

\newcommand{\blfootnote}[1]{%
  \begingroup
  \renewcommand{\thefootnote}{}%
  \footnotetext[0]{#1}%
  \endgroup
}

\title{The Covering Radius of $\RM(1,9)$ is Precisely $242$}

\author{Xiang-Dong Hou}
\author{Evelyn Sabo}
\author{Freeman Slaughter}
\affil{University of South Florida, Tampa FL, USA \\ \texttt{\{xhou, esabo, fslaughter\}@usf.edu} }
\date{\today}

\begin{document}
\maketitle
\vspace{-1.5em}

Let $\text{RM}(r,n)$ be the $r$th order Reed-Mullet code of length $2^n$, and let $\rho(r,n)$ denote its covering radius. It is known that $242\le\rho(1,9)\le 244$, and the determination of $\rho(1,9)$ is a long-standing open question.
We give a computer-assisted proof that $\rho(1,9)=242$, leveraging the classification of the $349$ orbits in $\RM(3,9)/\RM(2,9)$ under $\GL(9,2)$ \cite{BL,catalog}. 
The notebook to verify the computer search portion, \texttt{Verify\_RM(1,9).ipynb}, contains all data and recomputes each check, and may be found in the repository \url{https://github.com/FreemanSlaughter/Covering-Radius}. 
\blfootnote{This research was partially supported by NSF Grant DMS-2342254.}

\section{Preliminaries}

For $f:\F^n\to\F$, put $s_f(x)=(-1)^{f(x)}$ and
\[
 W_f(a)=\sum_{x\in\F^n}s_f(x)(-1)^{a\cdot x}.
\]
The character matrix $H_n$ is symmetric and satisfies $H_n^2=2^nI$.
Consequently
\begin{equation}\label{eq:walsh}
 \NL(f)=2^{n-1}-\tfrac12\M{W_f},\quad
 \sum_a W_f(a)^2=2^{2n},\quad
 \sum_aW_f(a)=2^ns_f(0).
\end{equation}
The first identity follows from
$d(f,a\cdot x+b)=2^{n-1}-(-1)^bW_f(a)/2$.
Thus $\rho(1,n)=\max_f\NL(f)$. At $n=9$, Parseval and the evenness of
Walsh coefficients give $\M{W_f}\ge24$, so $\rho(1,9)\le244$.
We will also use
\begin{equation}\label{eq:pairing}
 \sum_xs_f(x)W_p(x)=\sum_as_p(a)W_f(a),
\end{equation}
which is just the symmetry of $H_n$.
All polynomial degrees refer to algebraic normal form over $\F$.
A ``cubic'' below has degree at most three.

\begin{lemma}\label{lem:quotient}
If $W_f(a)\equiv c\pmod{2^k}$ for all $a$, where $1\le k<n$, then
$(W_f-c)/2^k\pmod2$ has degree at most $k-1$.
\end{lemma}
\begin{proof}
Regard the bits of $a$ as integer variables. The coefficient of $a^S$ in
the integer multilinear representation of $W_f$ is
\begin{equation}\label{eq:coefficient}
 (-2)^{|S|}\sum_{\substack{x_i=1 \\ i\in S}}s_f(x).
\end{equation}
When $|S|=k<n$, the inner sum has an even number of odd summands, so this
coefficient is divisible by $2^{k+1}$. The same divisibility is immediate
for $|S|>k$. Since $W_f-c$ is pointwise divisible by $2^k$, integer
M\"obius inversion shows that all its multilinear coefficients are
divisible by $2^k$. After division and reduction modulo two, every term of
degree at least $k$ vanishes.
\end{proof}

\begin{lemma}\label{lem:weight}
A nonzero Boolean polynomial of degree at most $d\le n$ has weight at
least $2^{n-d}$.
\end{lemma}
\begin{proof}
Choose a monomial of maximum degree $m \le d$. After any assignment of the
other $n-m$ variables, its coefficient is still one: no monomial properly
containing it occurs. Each of the $2^{n-m}$ disjoint restrictions is
therefore nonzero somewhere.
\end{proof}

\begin{lemma}\label{lem:cubic}
For a cubic on $m$ variables, every Walsh coefficient is divisible by
$2^{\lceil m/3\rceil}$. In particular, for a cubic $p$ on nine variables,
\begin{equation}\label{eq:residue}
 W_p(a)\equiv c+16\ell(a)\pmod{32}
\end{equation}
for some multiple $c$ of eight and some linear Boolean function $\ell$.
\end{lemma}
\begin{proof}
Include the character in the polynomial and expand its phase as a product
of factors $1-2x^S$, one for each nonconstant monomial. A term selecting
$j$ monomials whose union has size $u$ contributes, after summation, a
multiple of $2^{j+m-u}$. Since $u\le\min(m,3j)$, this exponent is at least
$\lceil m/3\rceil$. A constant term only changes the overall sign.

Applying this divisibility to the restrictions in
\eqref{eq:coefficient}, the coefficient indexed by $S$ for a nine-variable
cubic is divisible by $2^{|S|+\lceil(9-|S|)/3\rceil}$.
Linear coefficients are therefore multiples of $16$, and coefficients of
degree at least two are multiples of $32$. This gives \eqref{eq:residue}.
\end{proof}

\begin{lemma}\label{lem:quadratic}
Let $q$ be quadratic on $m$ variables. Its alternating bilinear form is
$B_q(u,v)=q(u+v)+q(u)+q(v)+q(0)$. If $r=\rank(B_q)$, then $r$ is even,
and its Walsh spectrum has $2^r$
nonzero coefficients, all of magnitude $2^{m-r/2}$. Hence
\[
 \sum_a|W_q(a)|=2^{m+r/2}.
\]
If $\wt(q)=2^{m-2}$, its support is an affine flat of codimension two.
\end{lemma}
\begin{proof}
Pairing off nonzero entries of the alternating form gives coordinates in
which
$q=x_1x_2+\cdots+x_{r-1}x_r+\lambda+\varepsilon$, with $\lambda$ linear.
Each paired character sum has magnitude two. Each unpaired coordinate
contributes either zero or two, and is nonzero for exactly one choice of
its frequency. This proves the spectral assertions.
If $\wt(q)=2^{m-2}$, then $W_q(0)=2^{m-1}$, forcing $r=2$.
There is no nonzero linear term on the radical, and translations in the
paired coordinates put $q$ in the form $x_1x_2$ or $x_1x_2+1$.
The specified weight selects the former. Its support is the intersection
of two independent affine hyperplanes.
\end{proof}

\section{Two Bounds on Cubic Spectra}

For a cubic $p$ on $\V$, define
\[
 L(p)=\sum_a|W_p(a)|,\quad
 C(p)=3W_p(0)+\sum_{a\ne0}|W_p(a)|.
\]
\begin{proposition}\label{prop:bounds}
Every such cubic satisfies $L(p)\le10240$ and $C(p)\le10368$.
\end{proposition}
\begin{proof}
The following elementary inequality will be used throughout, where we use $w_a$ to denote $W_p(a)$. 
If
$P_a(w_a)\ge|w_a|$, then
\begin{equation}\label{eq:majorant}
 3w_0+\sum_{a\ne0}|w_a|
 \le \sum_aP_a(w_a)+3w_0-P_0(w_0).
\end{equation}
Indeed, the difference between the right side and the left side is
$\sum\limits_{a\ne0}(P_a(w_a)-|w_a|)\ge0$.

First suppose the common residue modulo $16$ is eight.
Write $\ell(a)=a\cdot u$ in \eqref{eq:residue} and replace $p(x)$ by
$p(x+u)$. This multiplies $W_p(a)$ by $(-1)^{\ell(a)}$ and makes its
residue constant modulo $32$, since $-(c+16)\equiv c\pmod{32}$ for
$c\equiv8$ or $24$. Neither $L(p)$ nor $W_p(0)$ changes.
On the resulting lattice use
\[
\begin{array}{c|c|c}
 w\bmod32 & P(w)\ge |w| & \max(3w-P(w))\\ \hline
 8 & 3w^2/128-w/8+15/2 & 96\\
 24& 3w^2/128+w/8+15/2 & 80
\end{array}
\]
The maxima are over the indicated integer lattices.
For the first row, $P(w)-w=3(w-8)(w-40)/128$ when $w\ge0$,
and $P(w)+w=(w+24)(3w+40)/128$ when $w<0$.
Both are nonnegative on that lattice. Reflection proves the second row.
The two concave correction quadratics attain their lattice maxima at
$72$ and $56$, respectively. By \eqref{eq:walsh},
\[
 \sum_aP(W_p(a))=9984\mp\tfrac18\sum_aW_p(a)\le10048.
\]
Thus \eqref{eq:majorant} gives
\begin{equation}\label{eq:strict}
 L(p)\le10048<10240,\quad C(p)\le10048+96=10144.
\end{equation}

Otherwise all coefficients are divisible by $16$. Set
$t(a)=W_p(a)/16$ and $\epsilon(a)=t(a)\pmod2$.
Then $\epsilon$ is affine by \eqref{eq:residue}, and
$\sum_at(a)^2=1024$. The following bounds are immediate for even and odd
integers:
\[
\begin{array}{c|c|c|c}
 \epsilon & Q_a(t)\ge|t| & \sum_aQ_a(t(a))
             & 3t(0)-Q_0(t(0))\text{ is at most}\\ \hline
 0 & t^2/2 &512&4\\
 1 &(t^2+3)/4&640&8\\
 \text{nonconstant}&(t^2+\epsilon(a))/2&640&4
\end{array}
\]
In the last row $\wt(\epsilon)=256$.
Multiplying the last two columns' sum by $16$ in
\eqref{eq:majorant} bounds $C(p)$ by $8256$, $10368$, and $10304$,
respectively. The bounds on $L(p)$ are $8192$, $10240$, and $10240$.
\end{proof}

\section{Reduction to Two Cubic Spectra}

\begin{proposition}\label{prop:243}
The existence of a nine-variable function of nonlinearity $243$ would
imply the existence of one of nonlinearity $244$.
\end{proposition}
\begin{proof}
Suppose $\NL(f)=243$, so $\M{W_f}=26$.
For $n\ge2$ all Walsh coefficients are congruent to $-2\wt(f)$ modulo
four. Thus $f$ has odd weight. Let
$v=\bigoplus\limits_{x:f(x)=1}x$. Comparing the character sums gives
\[
 W_f(a)\equiv W_f(0)+4(a\cdot v)\pmod8.
\]
Translation by $v$ makes this residue constant: for odd-weight spectra
$c\equiv2$ or $6\pmod8$, and $-(c+4)\equiv c\pmod8$.
Taking the complement if necessary then gives $W_f\equiv2\pmod8$.
These operations preserve nonlinearity.

Put $s=s_f(0)$ and $z=(W_f-2)/8$. Then $-3\le z\le3$, and
\begin{equation}\label{eq:z}
 \sum_a z(a)=64s-128,\quad
 \sum_a z(a)^2=4128-32s.
\end{equation}
By Lemma~\ref{lem:quotient}, $q=z\pmod2$ is quadratic.
If $q\ne1$, Lemma~\ref{lem:weight} applied to $q+1$ gives
$\wt(q)\le384$. Since $z^2\le4+5q$, this would imply
$\sum\limits_{z} z^2\le4\cdot512+5\cdot384=3968$, contrary to \eqref{eq:z}.
Therefore it must be that $z\in\{-3,-1,1,3\}$ and
\[
 W_f\in\{-22,-6,10,26\}.
\]
If $s=1$, changing $f(0)$ from zero to one subtracts two from every
Walsh coefficient. The new spectrum lies in $\{-24,-8,8,24\}$, so the
new function has nonlinearity $244$ by Parseval.

If $s=-1$, then $\sum z^2=4160$, so exactly $56$ values have magnitude
one. Lemma~\ref{lem:quotient} makes
$p=(W_f-10)/16\pmod2$ cubic. The product $s_pz$ is one at those $56$
positions and minus three elsewhere. Equation~\eqref{eq:pairing} gives
\[
 \sum_xs_f(x)W_p(x)=\sum_as_p(a)W_f(a)=2W_p(0)-10496.
\]
Since $s_f(0)=-1$, it follows that
\[
 3W_p(0)-10496=\sum_{x\ne0}s_f(x)W_p(x)
       \ge-\sum_{x\ne0}|W_p(x)|.
\]
Thus $10496\le C(p)\le10368$, a contradiction.
\end{proof}

\begin{proposition}\label{prop:244}
Suppose $\M{W_f}\le24$ for a nine-variable Boolean function $f$.
Then $W_f\in\{-24,-8,8,24\}$, with $448$ coefficients of magnitude
$24$ and $64$ of magnitude eight. There is a cubic $p$ with
\begin{equation}\label{eq:equality}
 L(p)=10240,\quad s_f(x)=-\sgn W_p(x)\quad\text{when }W_p(x)\ne0,
\end{equation}
and its spectrum has one of the following forms:

\smallskip\noindent
\textup{I.} Every coefficient belongs to $\{-48,-16,16,48\}$.
There are $64$ coefficients of magnitude $48$ and $448$ of magnitude $16$.

\smallskip\noindent
\textup{II.} The parity of $W_p/16$ is nonconstant affine.
Its $256$ odd positions have values $\pm16$. The other positions contain
$192$ values $\pm32$ and $64$ zeros.
\end{proposition}
\begin{proof}
Odd weight would force every magnitude to be at most $22$, contradicting
Parseval. Hence $4\mid W_f$, and $W_f/4\pmod2$ is affine by
Lemma~\ref{lem:quotient}. If constantly one, all magnitudes are at most
$20$. If nonconstant, the energy is at most
$256(20^2+24^2)=249856<262144$. Therefore it is zero, and $8\mid W_f$.
Now $W_f/8\pmod2$ is quadratic. Unless constantly one, at least $128$
positions are divisible by $16$, so the energy is at most
$128\cdot16^2+384\cdot24^2=253952<262144$.
Thus $W_f\in\{-24,-8,8,24\}$, and Parseval gives the stated counts.

Set $p=(W_f-8)/16\pmod2$, which is cubic by
Lemma~\ref{lem:quotient}. The product $s_pW_f$ equals $-24$ on the $448$
large coefficients and eight on the others. Hence
\[
 -10240=\sum_xs_f(x)W_p(x)\ge-L(p)\ge-10240.
\]
Equality throughout proves \eqref{eq:equality}.
The strict bound \eqref{eq:strict} excludes residue eight modulo $16$,
and the constant-even case in Proposition~\ref{prop:bounds} is also
strict. In the constant-odd case, equality in
$|t|\le(t^2+3)/4$ forces $|t|\in\{1,3\}$, giving I.
In the nonconstant case, equality in
$|t|\le(t^2+\epsilon)/2$ forces $t=\pm1$ at odd positions and
$t\in\{0,\pm2\}$ at even positions, giving II.
In each case $\sum t^2=1024$ supplies the counts.
\end{proof}

\section{Two Exclusions}

We may change coordinates in $p$ and add affine terms without changing the
nonlinearity of its associated sign function. More explicitly, if
$p'(x)=p(Mx+u)+b\cdot x+\varepsilon$, with $M$ invertible, then
\begin{equation}\label{eq:equivalence}
 W_{p'}(a)=(-1)^{\varepsilon+(M^{-T}(a+b))\cdot u}
                 W_p(M^{-T}(a+b)).
\end{equation}
Thus the associated $f$ is changed by an affine input map and an affine
output term. The same statement applies to \eqref{eq:equality} when
zeros are present, by transforming the whole function $f$.

\begin{lemma}\label{lem:caseII}
Case II of Proposition~\ref{prop:244} is impossible.
\end{lemma}
\begin{proof}
By a linear change of variables and addition of a linear term to $p$, put
the odd frequency sector at $\alpha=1$. Write
$p(t,y)=p_0(y)+tq(y)$, where $y\in\F^8$ and $q$ is quadratic. Then
\[
 W_p(1,b)=2H_8(s_{p_0}\one_{\{q=1\}})(b),\quad |W_p(1,b)|=16.
\]
Parseval gives $\wt(q)=64$. Lemma~\ref{lem:quadratic} therefore makes
$\supp(q)=y_0+U$, with $\dim(U)=6$.
Choose an injective linear map $B:\F^6\to\F^8$ with image $U$, and set
$g(z)=p_0(y_0+Bz)$. The displayed formula becomes
\[
 W_p(1,b)=2(-1)^{b\cdot y_0}W_g(B^Tb).
\]
Since $B^T$ is onto, all coefficients of $W_g$ have magnitude eight.
Write $W_g=8s_{g^*}$. From $H_6^2=64I$ we obtain
$H_6s_{g^*}=8s_g$, so $g^*$ is bent as well.

The sign of $W_p(1,b)$ is therefore
$(-1)^{b\cdot y_0}s_{g^*}(B^Tb)$.
A six-variable function lifted along a surjective map from eight
variables has its nonzero Walsh coefficients multiplied by four.
Indeed, summing over each fiber gives zero unless the character
annihilates the two-dimensional kernel, and gives four otherwise.
Consequently this sign function has Walsh maximum $4\cdot8=32$.
By \eqref{eq:equality}, so does the restriction $f(1,\cdot)$.
But
\[
 W_{f(1,\cdot)}(b)=\frac{W_f(0,b)-W_f(1,b)}2,
\]
whose magnitude is at most $24$. This is a contradiction.
\end{proof}

Write $D_uh(x)=h(x+u)+h(x)$. For a homogeneous cubic $c$, define
\[
 D(c)=\{(D_uc)_2 \mid u\in\V\},
\]
where the subscript denotes the homogeneous quadratic part. This part is
$\sum\limits_i u_i(D_{e_i}c)_2$, so the map is linear.
Call $c$ nondegenerate when $\dim(D(c))=9$.

\begin{lemma}\label{lem:degenerate}
The cubic part of a polynomial arising in Case I of
Proposition~\ref{prop:244} is nondegenerate.
\end{lemma}
\begin{proof}
Otherwise $D_up$ is affine for some $u\ne0$.
If this derivative is constant, half the Walsh coefficients of $p$ vanish,
contrary to I. If nonconstant, it is invariant under translation by $u$.
Affine coordinates therefore put it at the coordinate $y$, with $u$ in
the $t$ direction, and give
\[
 p(t,y,z)=ty+r(y,z),\quad z\in\F^7.
\]
Summing first over $t$ yields
\[
 W_p(\alpha,\beta,\gamma)
   =2(-1)^{\alpha\beta}W_{r(\alpha,\cdot)}(\gamma).
\]
These coefficients are nonzero. Let $s_{h_\alpha}$ be the sign of
$W_{r(\alpha,\cdot)}$. Equation~\eqref{eq:equality} then gives
\[
 s_f(\alpha,\beta,\gamma)=-(-1)^{\alpha\beta}s_{h_\alpha}(\gamma),
 \quad W_f(a,b,d)=-2(-1)^{ab}W_{h_b}(d).
\]
Both seven-variable functions $h_b$ would consequently have Walsh
maximum at most $12$.

To exclude this, suppose $\M{W_h}\le12$ on seven variables.
Odd weight would bound all magnitudes by ten, violating Parseval.
Thus $4\mid W_h$ and $W_h/4\pmod2$ is affine.
Constant zero would bound all magnitudes by eight. Nonconstant parity
would give energy at most $64(12^2+8^2)=13312<16384$.
The parity is therefore constantly one. Parseval forces $112$ values of
magnitude $12$ and $16$ of magnitude four.
Now $k=(W_h-4)/8\pmod2$ is quadratic by Lemma~\ref{lem:quotient}, and
\[
 \sum_xs_h(x)W_k(x)=\sum_as_k(a)W_h(a)
   =-112\cdot12+16\cdot4=-1280.
\]
Lemma~\ref{lem:quadratic} bounds $\sum|W_k|$ by
$2^{7+6/2}=1024$, a contradiction.
\end{proof}

\section{Computation}

\subsection{Quadratic Portions to Check}

The published orbit count is, by \cite{BL,catalog},
\begin{equation}\label{eq:count}
 \#\bigl((\RM(3,9)/\RM(2,9))/\GL(9,2)\bigr)=349.
\end{equation}
We use the $349$ polynomials recorded in Langevin's catalog \cite{catalog}.
The audit below verifies that they are pairwise inequivalent, so
\eqref{eq:count} makes the list complete.
After a linear change of coordinates, any cubic is $c+q+\lambda$, with
$c$ from this list, $q$ homogeneous quadratic, and $\lambda$ affine.
By \eqref{eq:equivalence}, it suffices to consider $p=c+q$.
There are $\binom92=36$ coefficients in $q$.

\begin{lemma}\label{lem:finite-reduction}
The common residue of $W_{c+q}$ modulo $16$ depends only on $c$.
If it is zero, the map
\begin{equation}\label{eq:phi}
 \Phi_c(q)_i=\frac{W_{c+q}(e_i)-W_{c+q}(0)}{16}\pmod2,
 \quad i=1,\ldots,9,
\end{equation}
is affine in the $36$ coefficients of $q$.
Its vanishing means that $W_{c+q}$ is constant modulo $32$.
If the system is consistent, write its solution space as $q_0+K_c$.
Then $D(c)\subseteq K_c$.
One representative of each coset modulo $D(c)$ suffices for Case I.
\end{lemma}
\begin{proof}
In the nine-variable phase expansion from Lemma~\ref{lem:cubic}, a term
can survive modulo $16$ only when $j+9-u\le3$. Since
$u\le\min(9,3j)$, this forces $j=3$ and $u=9$.
Thus only three disjoint cubic monomials can contribute. Quadratic and
frequency-linear monomials do not affect the residue.

For \eqref{eq:phi}, use
\[
 W_{c+q}(e_i)-W_{c+q}(0)
   =-2\sum_{x_i=1}(-1)^{c(x)+q(x)}.
\]
Compute the eight-variable sum modulo $16$.
The same expansion can retain only $j=3$, $u=8$.
The possible degree triples are $(3,3,3)$ and $(3,3,2)$, so a surviving
term involves at most one quadratic coefficient. Cubic coefficients are
fixed, and quadratic coefficients in the restriction are affine in $q$.
A possible constant sign makes no difference modulo $16$, since the
retained terms are multiples of eight. Dividing the displayed difference
by $16$ and reducing modulo two proves affinity.
Equation~\eqref{eq:residue} shows that $\Phi_c$ records the linear part
of the affine function $W_{c+q}/16\pmod2$. This proves the vanishing
criterion.

Translation by $u$ changes the quadratic part of $c+q$ by $(D_uc)_2$,
with the remaining change affine. In a spectrum divisible by $16$,
multiplication by a sign has no effect modulo $32$.
An affine term added to the polynomial shifts its frequency argument and
possibly negates its spectrum. These operations leave the linear part of
$W_{c+q}/16\pmod2$ unchanged. Thus $D(c)$ lies in the homogeneous kernel
$K_c$. Equation~\eqref{eq:equivalence} also shows that each such coset
has a fixed spectral magnitude multiset and fixed sign-function
nonlinearity. Choosing a complement to $D(c)$ in $K_c$ is sufficient.
\end{proof}

The code obtains the affine map by evaluating it at zero and the $36$
quadratic monomials. It solves the equations by Gaussian elimination,
then extends a basis of $D(c)$ to a basis of $K_c$.
The $124$ residue-eight forms are excluded by \eqref{eq:strict}.
Of the other $225$, exactly $32$ are degenerate and are excluded by
Lemma~\ref{lem:degenerate}. All $193$ remaining systems are consistent.
For the complements $B_c$, the computed dimensions are
\[
\begin{array}{c|rrrrrrr}
 \dim(B_c) &18&19&20&21&22&24&27\\ \hline
 \text{number of classes}&142&20&18&6&5&1&1
\end{array}
\]
Hence $\sum\limits_c2^{\dim(B_c)}=251133952$.

\subsection{Completeness}

The audit uses contraction and restriction invariants as in~\cite{BL}.
Let $T(u,v,w)=D_uD_vD_wc(0)$ be the alternating trilinear form of $c$.
The correspondence between cubic parts and such forms is
$\GL(9,2)$-equivariant: $T(e_i,e_j,e_k)$ is the coefficient of
$x_ix_jx_k$. Put $r(u)=\rank(T(u,\cdot,\cdot))$ and
$\psi_j(u)=\one_{\{r(u)=2j\}}$ for $0\le j\le4$.
Here a hat denotes the unnormalized Fourier transform of an integer-valued
function, with the natural pairing between a vector space and its dual.

For $H=a^\perp$, $a\ne0$, let $r_H(u)$ be the contraction rank for
$T|_H$, and $\chi_{H,j}(u)=\one_{\{r_H(u)=2j\}}$, $0\le j\le3$.
These are all possible ranks: each nonzero $u\in H$ belongs to the
radical of its contraction, and $\dim(H)=8$.
Define the multiset
\[
 \Gamma(H)=\bigl\{(\widehat\chi_{H,0}(b),\ldots,
                    \widehat\chi_{H,3}(b)) \mid b\in H^*\bigr\}.
\]
An isomorphism of $H$ merely permutes its dual, so $\Gamma(H)$ is
basis-independent. Assign common labels to the distinct exact multisets
occurring anywhere in the catalog. There are $32$ of them.
For each label $j$, let $\phi_j(a)$ indicate that label at $a\ne0$,
and set $\phi_j(0)=0$. The audit compares the following two multisets
jointly:
\begin{align*}
 &\bigl\{(\Gamma(a^\perp),\widehat\psi_0(a),\ldots,
                      \widehat\psi_4(a)) \mid a\ne0\bigr\},\\
 &\bigl\{(r(u),\widehat\phi_0(u),\ldots,
                      \widehat\phi_{31}(u)) \mid u\in\V\bigr\}.
\end{align*}
They are invariants. Under $T'(u,v,w)=T(Mu,Mv,Mw)$, one has
$r'(u)=r(Mu)$ and
$\widehat\psi'_j(a)=\widehat\psi_j(M^{-T}a)$.
Also $\Gamma'(a^\perp)=\Gamma((M^{-T}a)^\perp)$, so
$\widehat\phi'_j(u)=\widehat\phi_j(Mu)$.
Each displayed multiset is therefore only permuted.

For efficiency the code uses, for $u\in a^\perp$,
\[
 r_{a^\perp}(u)=
 \begin{cases}
 r(u)-2,&a\in\operatorname{im}T(u,\cdot,\cdot),\\
 r(u),&a\notin\operatorname{im}T(u,\cdot,\cdot).
 \end{cases}
\]
To justify this, let $R$ be the radical of the contraction. If
$a\in\operatorname{im}T(u,\cdot,\cdot)=R^\perp$, then $a^\perp$
contains $R$ and descends to a hyperplane in the symplectic quotient,
reducing rank by two. Otherwise $a^\perp$ maps onto that quotient and
retains its rank.

The code sorts and compares full integer tuples, preserving
multiplicities, and obtains $349$ distinct pairs. 
Combined with \eqref{eq:count}, this proves that each orbit is
represented. 
The checksum checks the transcription only. 

\subsection{Exhaustive Check}

For every $q\in q_0+B_c$, the code computes $W_{c+q}$.
It first checks $W_{c+q}(0)\equiv16\pmod{32}$, then retains precisely
those spectra in $\{-48,-16,16,48\}^{512}$.
For each retained spectrum it forms $s_f=-\sgn W_{c+q}$ and computes
$H_9s_f$. The completed computation gives
\begin{equation}\label{eq:computed}
 \begin{aligned}
 \text{quadratic representatives tested}&=251133952,\\
 \text{spectra retained}&=90126659,\\
 \max(\NL(f))\text{ among retained spectra}&=240.
 \end{aligned}
\end{equation}

Two preparations are used: row elimination with full Walsh transforms,
and reverse column elimination with direct character sums at
$\{0,e_1,\ldots,e_9\}$. They choose different offsets and complementary
bases. The first enumeration stores phases in signed $16$-bit arrays
and uses ascending Walsh stages. The second stores truth bits, uses
$32$-bit arrays and descending stages, and checks every transformed
coefficient's residue. Their tested counts, retained counts, and maximum
nonlinearities agree separately for all $193$ classes.
Every transform intermediate has magnitude at most $512$, so these
integer types suffice. The code also compares both transforms with
direct character sums on fixed test tables.

Each complement is traversed in binary reflected Gray order
$g(i)=i\mathbin{\mathtt{xor}}\lfloor i/2\rfloor$.
This is a bijection on the coefficient vectors. Between consecutive
indices, only the bit at the least significant one of $i$ changes.
Parallel chunks begin at their exact Gray vector and partition the full
interval $0\le i<2^{\dim(B_c)}$ without omissions.
Thus \eqref{eq:computed} is an exhaustive integer computation. 
In view of Lemma~\ref{lem:finite-reduction}, it excludes Case I
of Proposition~\ref{prop:244}.

\section{Reproduction}

Kavut and Y\"ucel \cite{KY} give the following truth table in hexadecimal
\cite[\S3.1]{KY}:
\begin{center}\footnotesize\ttfamily
68B7EF2DA03B0D3EA00DB6A96DD99AEAFDB9C842B6D5DC8C4526CE0DD29020DB\\
B75FE3314568344E73688FF0CB2482E065231869E1AA4583765CC491F8A8DB12
\end{center}
Concatenate the lines, expand each digit most significant bit first, and
index the resulting bits by the integers $\{0,\ldots,511\}$.
All $512$ direct character sums are checked by the Python file, giving
Walsh maximum $28$ and hence nonlinearity $242$.

\begin{theorem}
$\rho(1,9)=242$.
\end{theorem}
\begin{proof}
Equation~\eqref{eq:walsh} gives the upper bound $244$.
If nonlinearity $244$ were attained, Proposition~\ref{prop:244} would
produce Case I or II. Lemma~\ref{lem:caseII} excludes II.
Lemma~\ref{lem:degenerate}, Lemma~\ref{lem:finite-reduction}, the catalog
audit and the computation \eqref{eq:computed} exclude I.
Proposition~\ref{prop:243} then excludes $243$.
The witness from Kavut and Y\"ucel attains $242$.
\end{proof}

Run the accompanied file \texttt{Verify\_RM(1,9).ipynb}, which can be found at \url{https://github.com/FreemanSlaughter/Covering-Radius}, in a Jupyter Notebook such as Google Colab to verify.
The script includes the catalog of \texttt{classification.txt} from \cite{catalog}, and needs no additional inputs or network access. 
It performs both full enumerations, compares every class, and generates \nolinkurl{rm19_verification.json}. 
This file is included in the repository above. 

\section{Acknowledgments and Disclosure}

This research was partially supported by NSF Grant DMS-2342254.

We acknowledge that discussions with AI were helpful in formulating and refining some of these proofs.
Nonetheless, we the authors take full responsibility for the accuracy of the results contained in this manuscript.
We aim to publish a more complete, understandable, and streamlined version of this document soon.

\bibliographystyle{splncs04}
\bibliography{archive.bib} 
\end{document}